# Deteriorated Interlayer Coupling in Twisted Bilayer Cobaltites

*Dongke Rong,* [†,‡,◊] *Xiuqi Chen,* [†,‡,◊] *Shengru Chen,* [†,‡,∆,◊] *Jinfeng Zhang,* [§] *Yue Xu,* [†,‡], *Yanxing Shang,* [†,‡] *Haitao Hong,* [†,‡] *Ting Cui,* [†,‡] *Qianying Wang,* [†,‡] *Chen Ge,* [†,‡] *Can Wang,* [†,‡] *Qiang Zheng,* [‖] *Qinghua Zhang,* [†,‡] *Lingfei Wang,* [§] *Yu Deng,* [⊥,*] *Kuijuan Jin,* [†,‡,*] *Gang-Qin Liu,* [†,‡,*] *and Er-Jia Guo* [†,‡,*]

**Affiliations:**

† Beijing National Laboratory for Condensed Matter Physics and Institute of Physics, Chinese Academy of Sciences, Beijing 100190, China

‡ Department of Physics & Center of Materials Science and Optoelectronics Engineering, University of Chinese Academy of Sciences, Beijing 100049, China

∆ School of Physics, Zhejiang University, Hangzhou 310058, China

§ Hefei National Research Center for Physical Sciences at Microscale, University of Science and Technology of China, Hefei 230026, China

‖ CAS Key Laboratory of Standardization and Measurement for Nanotechnology, CAS Center for Excellence in Nanoscience, National Centre for Nanoscience and Technology, Beijing 100190, China

⊥ National Laboratory of Solid-state Microstructures, Jiangsu Key Laboratory of Artificial Functional Materials, College of Engineering and Applied Sciences, Collaborative Innovation Center of Advanced Microstructures, Nanjing 210023, China

**KEYWORDS:** twistronics, magnetic coupling, freestanding membranes, moiré pattern

## ABSTRACT

A wealth of remarkable behaviors is observed at the interfaces between magnetic oxides due to the coexistence of Coulomb repulsion and interatomic exchange interactions. While previous research has focused on bonded oxide heterointerfaces, studies on magnetism in van der Waals interfaces remain rare. In this study, we stacked two freestanding cobaltites with precisely controlled twist angles. Scanning transmission electron microscopy revealed clear and ordered moiré patterns, which exhibit an inverse relationship with the twist angle. We observed that the Curie temperature in the twisted region is lower than that in the single-layer region and varies systematically with the twist angle. This phenomenon may be related to the weakening of the orbital hybridization between oxygen ions and transition metal ions in the unbonded interfaces. Our findings suggest a potential avenue for modulating magnetic interactions in correlated systems through twist, providing opportunities for the discovery of unknown quantum states.

## MAIN TEXT

Magnetic coupling at interfaces plays a pivotal role in the functionality of spintronic devices. For example, interlayer RKKY interactions and exchange bias effects are essential for stabilizing the magnetic reference layer in magnetic tunnel junctions (MTJs).[1-5] Meanwhile, interlayer magnetic dipole coupling and chiral interactions hold promise for the development of novel computing devices, such as magnetic logic circuits and neuromorphic computing units.[6-9] Current research on interlayer magnetic coupling predominantly focuses on chemically bonded heterointerfaces between magnetic or non-magnetic materials, including complex metal oxide heterostructures.[10-16] These interfaces, formed through covalent or ionic bonds, exhibit a diverse array of novel magnetic phenomena distinct from those of the parent materials. Observed interfacial phenomena, such as charge transfer, spin-orbit coupling, and lattice structure regulation, arise from the intricate interactions between transition metal ions mediated by anions.

Beyond conventional bonded magnetic interfaces, van der Waals (vdW) magnetic interfaces and their associated physical properties remain largely unexplored. The atomic distance between the magnetic layers in these systems is larger than that within the layers, leading to complex magnetic interactions that depend on factors such as the number of atomic layers and the stacking order of the material.[17, 18] Although numerous studies have demonstrated intriguing magnetic coupling in two-dimensional (2D) magnetic materials like $Cr_2Ge_2Te_6$, $CrI_3$, and CrSBr, among others,[19-26] vdW magnetic interfaces in complex oxides have been scarcely investigated. Recent advances have introduced new methods for fabricating freestanding oxide membranes.[27, 28] These membranes, akin to 2D materials, offer extreme flexibility and can be tailored by controlling stacking sequences and twist angles between subsequent layers. Unlike traditional oxide heterostructures, stacking freestanding oxide membranes does not require lattice matching or the formation of chemical bonds, allowing layers with different lattice symmetries and atomic spacings to be combined. Twisting these membranes can generate moiré patterns, similar to those seen in twisted 2D materials, opening up new avenues for exploring unique atomic structures and physical properties.[29-31] While previous studies have focused on the atomic arrangements in twisted oxide membranes with controllable twist angles,[32, 33] the exploration of artificially twisted stacking in vdW magnetic oxide membranes remains unexplored.

In this study, we fabricated high-quality, single-crystalline $La_{0.8}Sr_{0.2}CoO_3$ (LSCO) freestanding membranes using a water-soluble sacrificial layer. Distinct and periodic moiré patterns were observed in the twisted LSCO bilayer membranes, with the spacing and size of the moiré patterns being inversely proportional to the twist angle between the crystallographic axes of the constituent layers. Nanoscale magnetic probing revealed a reduced magnetic transition temperature in the twisted region, indicating that twistronics offers a new degree of freedom for designing and engineering oxide interface phenomena.

The water-soluble sacrificial layer $Sr_4Al_2O_7$ (T-SAO) [28] and the magnetic LSCO layer were epitaxially grown on (001)-oriented $(LaAlO_3)_{0.3}$-$(Sr_2AlTaO_6)_{0.7}$ (LSAT) substrates using pulsed laser deposition (PLD) (see Methods). The structural properties of LSCO were characterized using an x-ray diffractometer (XRD) before release from the substrates. Both LSCO and T-SAO films exhibited distinct Laue oscillations, indicating extremely high crystallinity and successful epitaxial growth (Figure S1). The out-of-plane lattice constant ($c$) of as-grown

LSCO is approximately 3.805 Å, suggesting the LSCO films suffer an in-plane tensile strain. Following the conventional transfer method, we were able to attach an LSCO freestanding membrane onto a bare $Al_2O_3$ substrate. An XRD 2θ-ω scan revealed a single-crystalline LSCO membrane, which exhibited an increased $c$ to 3.835 Å due to the release of epitaxial tensile strain. Thus, the freestanding LSCO membranes are free of substrate's clamping induced epitaxial strain effects.

To obtain a substantial portion of LSCO freestanding membranes, we employed an alternative transfer technique that can be readily applied to various supports, as illustrated in Figure S2 and detailed see Methods. The structural properties of twisted BL LSCO were investigated by planar-view high-angle annular dark field (HAADF) scanning transmission electron microscopy (STEM) experiments (Figure 1a). We intentionally selected the sample edge area to include single layer (SL), BL, and multi-layer regions within a single region (Figure S3). For non-twisted BL LSCO, where the twist angle α is approximately 0°, the upper and lower LSCO lattices overlap in a highly coincidental (atom-on-atom) manner. We present a detailed simulation analysis of twisted oxide bilayers (Figure S4). The simulation results clearly demonstrate that the HAADF image and Moiré pattern can identify structural modifications in bilayers with a twist angle as small as 0.1°. Therefore, if the twist angle is not precisely 0°, it is likely below 0.1°, which is indistinguishable and therefore negligible in its influence. The color difference due to mass-thickness contrast was visible during STEM imaging, with darker regions corresponding to the SL LSCO membrane and brighter regions to the non-twisted BL LSCO membrane. These regions were separated by an atomically sharp boundary. The atomic arrangement of LSCO unit cells is depicted in Figure 1b, where La/Sr atoms occupy the A-site, Co atoms are located at the B-site, and O atoms form the octahedra. Figure 1c presents the line profiles of HAADF intensity across the SL and BL LSCO regions. The thickness of the upper LSCO membranes is approximately 5 nm, which is consistent with the thickness determined by PLD growth. The gap distance between twisted BL LSCO, depending on the transfer process, does not keep constant. We average the distance is approximately 2 ± 0.5 nm (Figure S5). We understand that the gap maybe a bit wide for a dipole-dipole interaction. Please note that the roughness and step-and-terrace of LSCO membranes is approximately 1-2 unit-cells. Thus, the magnetic coupling between twisted LSCO still valid.

For twisted BL LSCO, the characteristic moiré features were observed. Figure 1e presents a representative STEM image that includes both SL and twisted BL regions. Although the α was pre-set during the transfer process, the exact twist angle of the stacked freestanding membranes was determined in two independent methods. The first method involves measuring the twist angle between the two scattering vectors directly from the fast Fourier transform (FFT) image. The second method involves measuring the real space distance ($d_{\text{moiré}}$) of the moiré fringes and $d_{\text{vector}}$. The twist angle is then calculated using the formular α = 2 $\sin^{-1} d_{vector}/2d_{\text{moiré}}$. In Figure 1e, $d_{\text{vector}}$ = 1.925 Å for the (200) scattering vector and $d_{\text{moiré}}$ = 14.91 Å, resulting in α ~ 7.4°. The calculated twist angle is consistent with both the FFT result and the pre-set misalignment angle. We further investigated the structural properties of moiré patterns formed at a twist angle of α = 4°, 7.4°, 9°, 12°, 19°, and 39° twisted BL LSCO membranes. The STEM-HAADF images and their corresponding FFT results are summarized in Figures 2a and 2b, respectively. Periodic moiré patterns, marked by dashed yellow squares, are observed in all twisted BL LSCO membranes. The spacing and size of the moiré patterns progressively decrease with increasing twist angles. In the FFT images, two sets of diffraction spots, indicated by red and blue rectangles, confirm the overlapping of two LSCO freestanding membranes with a misorientation. We also simulated the moiré patterns and corresponding FFT results with different twist angles (Figure S6). Both experiments and simulations are consistent with each other. Similarly, we employed two calculation methods to determine and verify the twist angles of each BL LSCO. Specifically, Figure 2c presents a zoomed-in view of a representative single moiré pattern from a twisted BL LSCO with α = 19°. The bright white spots correspond to La/Sr atoms, while the Co atoms are less distinct, and the O atoms are invisible due to the HAADF intensity scaling with the atomic number of the elements.

To elucidate the atomic arrangement in the twisted BL LSCO, a 3 × 3 unit-cell model with α = 19° was constructed for direct comparison with our measured HAADF image (Figure 2d). The dependence of the moiré pattern area ($A^2$) and spacing (λ) on the twist angle is summarized in Figures 2e and 2f, respectively. Typically, the moiré effect occurs when two periodic layers are superimposed and the general relationship is $\lambda \sim \frac{(1+\delta)\alpha}{\sqrt{2(1+\delta)(1-\cos\alpha)+\delta^2}}$, where δ is the lattice mismatch between the two layers.[34,35] When two layers are made of same material with identical lattice parameters, the formula can be simplified to $\lambda \sim \frac{a}{2\sin\left(\frac{\alpha}{2}\right)}$. We find that both parameters in our twist BL LSCO follow the reciprocal relation to $\frac{a}{2\sin\left(\frac{\alpha}{2}\right)}$ and $\left(\frac{a}{2\sin\left(\frac{\alpha}{2}\right)}\right)^2$ for λ and $A^2$, respectively. The $\lambda \sim \frac{a}{2\sin\left(\frac{\alpha}{2}\right)}$ relationship agrees well with previously reported twist 2D materials, such as twist bilayer graphene (tBLG), twist graphene-hBN heterostructure and others transition metal dichalcogenides (TMDs).[19, 36-38] Early work demonstrated that a special critical twist angles can generate novel physics, such as flat band, superconductivity, and correlated insulating state, due to the strong correlated electron interaction. Although the crystal structure and symmetry of perovskite LSCO is different from the hexagonal lattice of graphene and TMDs, they all follow the same rule: $\lambda \sim \frac{a}{2\sin\left(\frac{\alpha}{2}\right)}$, confirming the universal moiré effect from a geometric point of view.

In addition to the moiré patterns observed in the twisted BL LSCO, it is noteworthy that moiré stripes are formed within the BL LSCO (Figure S7). These stripes have a width of approximately 2 nm and align at 45° with respect to both the (100) and (010) orientations of the SL LSCO. We ruled out the possibility of a tilted zone axis during the STEM imaging process. We meticulously simulated the tilt patterns of BL LSCO with various tilted zone axes (Figure S8). To generate a stripe with a width as narrow as 2 nm, the zone axis would need to be increased to at least 400 mrad, which clearly does not apply to our case. Thus, we attribute the observed phenomena to the lateral shift of the upper LSCO layer. The clear structural variation with the twist angle opens up new avenues for engineering correlated electron systems using "*moiré modulation*" as an important degree of freedom.

Our previous work demonstrated that freestanding membranes, when attached to one another, form a space layer with a thickness of at least 1 nm, even after being heated to the growth temperature (typically 800 °C).[39,40] This result suggests that no chemical bonds are formed between the two LSCO membranes; instead, vdW interactions may still occur at the interfaces. A recent study by Sánchez-Santolino *et al.* [33] reported that lateral strain modulation forms between twisted freestanding ferroelectric oxides due to interface matching. They found that the nanoscale-modulated distribution of shear strain drives notable polar vortices and antivortex structures, which appear to change with the twist angle. To our knowledge, this report is the first experimental evidence that vdW interactions between twisted membranes can lead to moiré-related phenomena. To study the magnetic interactions at vdW interfaces formed in twisted BL LSCO, we performed spatially resolved magnetic measurements using a home-built optically detected magnetic resonance (ODMR) system. As shown in Figure 3a, a layer of shallow nitrogen-vacancies (NV centers) in a high-purity diamond (Element Six, with an initial nitrogen concentration of 5 ppb) serves as *in-situ* quantum sensors. The NV centers were generated by nitrogen ion implantation with an energy of 10 keV and a dose of $5 \times 10^{13}$ ions/$cm^2$ and subsequent high-temperature annealing (1000 °C for 70 mins). The density of these NV centers was estimated to be approximate 60 ppb, with a spin coherence time $T_2$ = 1.31 (6) μs. The twisted BL LSCO with $\alpha = 8°$ were transferred to the diamond substrate before the measurements. After loading the sample, the regions of interest were addressed with both the brightfield optical image and the confocal laser scanning image.

Figure 3b shows the comparison of ODMR spectra from both SL and BL LSCO regions at different temperatures (more results are shown in Figure S9). At low temperatures, the ODMR spectra split with a magnitude of $2\gamma B$, where $\gamma$ = 2.802 MHz/Gauss is the gyromagnetic ratio of NV electron spin and $B$ is the magnetic stray field from the LSCO film. Figures 3e and 3f summarize the temperature-dependent ODMR splitting for SL and BL LSCO membranes, respectively. The ODMR splitting for BL LSCO increases nearly twice as much as that of SL LSCO, indicating that the net magnetic moment increases when the layer thickness doubled. We fit the $2\gamma B$-$T$ curve with the Curie-Weiss law. Interestingly, we found that the magnetic transition temperatures ($T_C$) are approximately 187 K and 200 K for BL and SL LSCO, respectively. These results were repeated at different positions in both SL and BL LSCO (Figure

S10), suggesting the robust reduction of $T_C$ after stacking two BL with a twist angle. Since the ensemble NV centers are measured in single laser spots, the gradient of the stray field also leads to a significant broadening of the ODMR spectra. Here, the ODMR width also shows a clear temperature dependence with a lower $T_C$ for the BL region (Figure S11). The $T_C$ of LSCO films is known to increase with film thickness.[41,42] Finite-size scaling is particularly evident in films with thicknesses below 40 nm. However, our findings indicate that the $T_C$ for BL LSCO is reduced compared to SL LSCO, which contrasts with the expected effects of doubling the film thickness. We attribute this observed reduction in $T_C$ to the magnetic coupling between the LSCO layers.

To further validate the observed phenomena, we fabricated BL LSCO samples with varying twist angles. We aimed to reduce the LSCO film thickness to approximately 3 nm (around 6–7 unit cells) for three key reasons: (1) to enhance magnetic interactions with larger interface volume, (2) to maintain a larger sample fraction, and (3) to confirm the robustness of the observed effects. Thus, the $T_C$ of these thinner samples is lower than that of the previous, thicker ones, consistent with earlier reports.[41,42] Figures 4a–4d show optical images of BL LSCO with twist angles of $\alpha$ = 2.8°, 5.1°, 15.4°, and 21.9°, respectively. The completeness of these thinner samples is superior to that of previous 5 nm-thick BL LSCO samples. Due to variations in sample size and quality, direct comparisons of absolute magnetization values and coercive fields are not meaningful. Figure 4e presents the temperature-dependent magnetization of these samples, revealing a clear trend: $T_C$ increases progressively with increasing α\alphaα. Figure 4f summarizes $T_C$ as a function of twist angle, showing a trend consistent with NV magnetometry results. This agreement strongly suggests that twisting plays a crucial role in tuning magnetic interactions in correlated oxides.

Since the major impact factors, such as epitaxial strain, Sr concentration, and film thickness,[43-45] are constant for BL and SL LSCO, leaving twist angle is a single parameter that affects the magnetic properties of BL LSCO. Previously, Xie *et al*. report a reduction of ~20 K of the Néel temperature is observed in the 1.1° twist 2D magnetic $CrI_3$, compared with natural $CrI_3$.[46] They attribute this to the moiré superlattice-induced magnetic competition. The emergence of non-collinear spins in the twisted $CrI_3$ modulates the interlayer coupling, leading to the reduction of critical temperature. Our result shows similar reduced $T_C$ in twisted BL LSCO, suggesting the weakening of magnetic intercoupling between two freestanding membranes. However, unfortunately, the exact spin structure and magnetic interaction in the twisted BL LSCO is much more complex than in 2D materials due to the correlated electron system. Previous research has shown that the ordering of Co-3*d* orbitals is responsible for the ferromagnetism observed in cobaltites. Consequently, one potential scenario is that the hypothesized reconstruction of orbital ordering, achieved by twisting two LSCO layers, could influence the magnetic exchange interactions, potentially leading to a reduction in the magnetic ordering temperature. Similar behaviors have been observed in other perovskite oxides, where the transition temperature of the orbital order increases while the magnetic order decreases.[47,48] Another possibility is the influence of phase separation at the twisted LSCO interfaces. This instability is promoted by a reduction in carriers, which weakens the ferromagnetic coupling between Co ions and enhances the relevance of superexchange antiferromagnetic interactions. Although further experimental and theoretical investigations are necessary to fully understand the microscopic nature of these complex electronic order phases at the interfaces, the current findings offer valuable insights into the critical surface/interface sensitivity of orbitally-active twisted LSCO.

For the twisted BL LSCO, we still need to understand exactly how the magnetic exchange interaction varies with the twist angle. It is certainly both interesting and important to map the evolution of magnetic properties—such as saturation moment, critical temperature, and anisotropy—as a function of twist angle. This is not merely a simple method for tuning physical properties; more importantly, it is crucial to understand how the coupling strength of interfacial magnetic exchange interaction changes under different stacking orders at the interface. Recent years, the scanning NV microscopy at cryogenic temperatures has also facilitated the study of magnetic domain structures and magnetic fluctuation around the Curie temperatures, especially in two-dimensional ferromagnetic or antiferromagnetic materials. It is not only a highly sensitive magnetic detection method but also possesses ultra-high spatial resolution, which can help us to investigate the evolution of magnetic domains in twisted and non-twisted regions under external fields, including coercivity, domain wall motions, magnetic reversal time, and so on. Furthermore, the development of new theoretical methods and computational tools is essential for understanding vdW correlated magnetic interfaces, which is also one of the key development directions in future.

In summary, we successfully fabricated twisted bilayer LSCO membranes with controllable twist angles. High-resolution scanning transmission electron microscopy revealed clear and ordered moiré patterns. The period and area of the moiré patterns exhibit a reciprocal relationship with the twist angle, which is consistent with the universal rule $\lambda \sim \frac{a}{2\sin\left(\frac{\alpha}{2}\right)}$ in twisted moiré structures. NV magnetometry measurements indicate a reduction in the Curie temperature by approximately 13 K in twisted BL LSCO compared to single-layer LSCO, suggesting weakened magnetic coupling at the van der Waals interface between the LSCO membranes. Our results suggest that the twist angle can serve as a novel degree of freedom for modulating magnetic interactions in correlated systems. Furthermore, we note that a recent study has demonstrated the formation of atomically clean interfaces during membrane stacking following high-temperature post-annealing in oxygen.[49] We believe that these structurally well-defined interfaces will facilitate moiré-type reconstruction and promote exciting phenomena.

## FIGURES AND FIGURE CAPTIONS

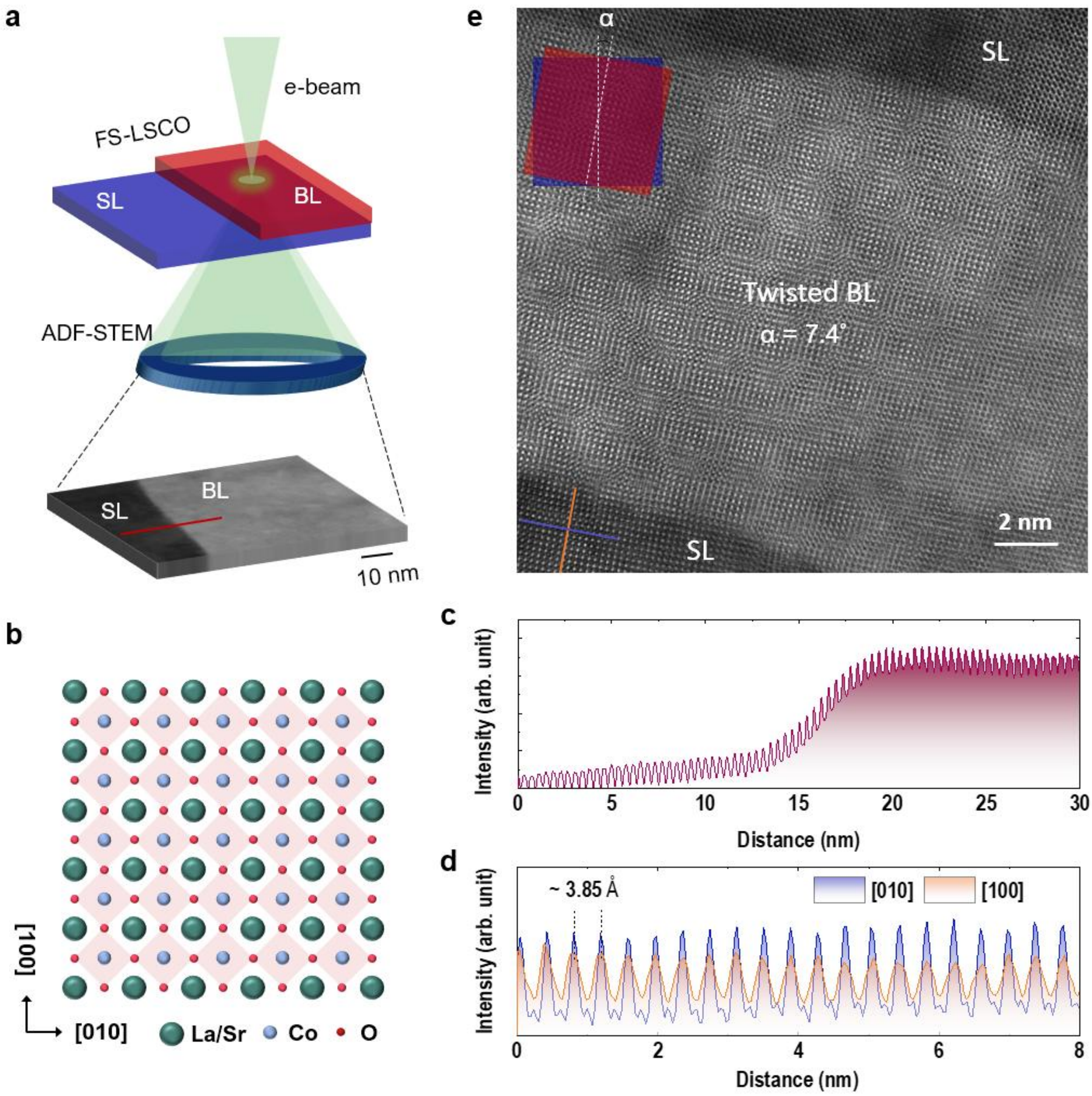

**Figure 1. Structural characterization of twisted BL LSCO.** (a) Schematic of atomic structure imaging via HAADF-STEM. (b) Schematic of a perovskite-type LSCO, exhibiting isotropy along (100) and (010) orientations. (c) Intensity profile obtained from line scans averaged across SL and BL regions in (a). (d) Averaged intensity profile lines extracted from (010) and (100) orientations at the SL region, revealing a lattice constant of 3.85 Å in both orientations, confirming the in-plane isotropy of the SL LSCO membranes. (e) High-resolution STEM image depicting the LSCO membranes with a single layer (SL) region and a twisted bilayer (BL) region at a rotation angle α = 7.4°. The white scale bar denotes 2 nm.

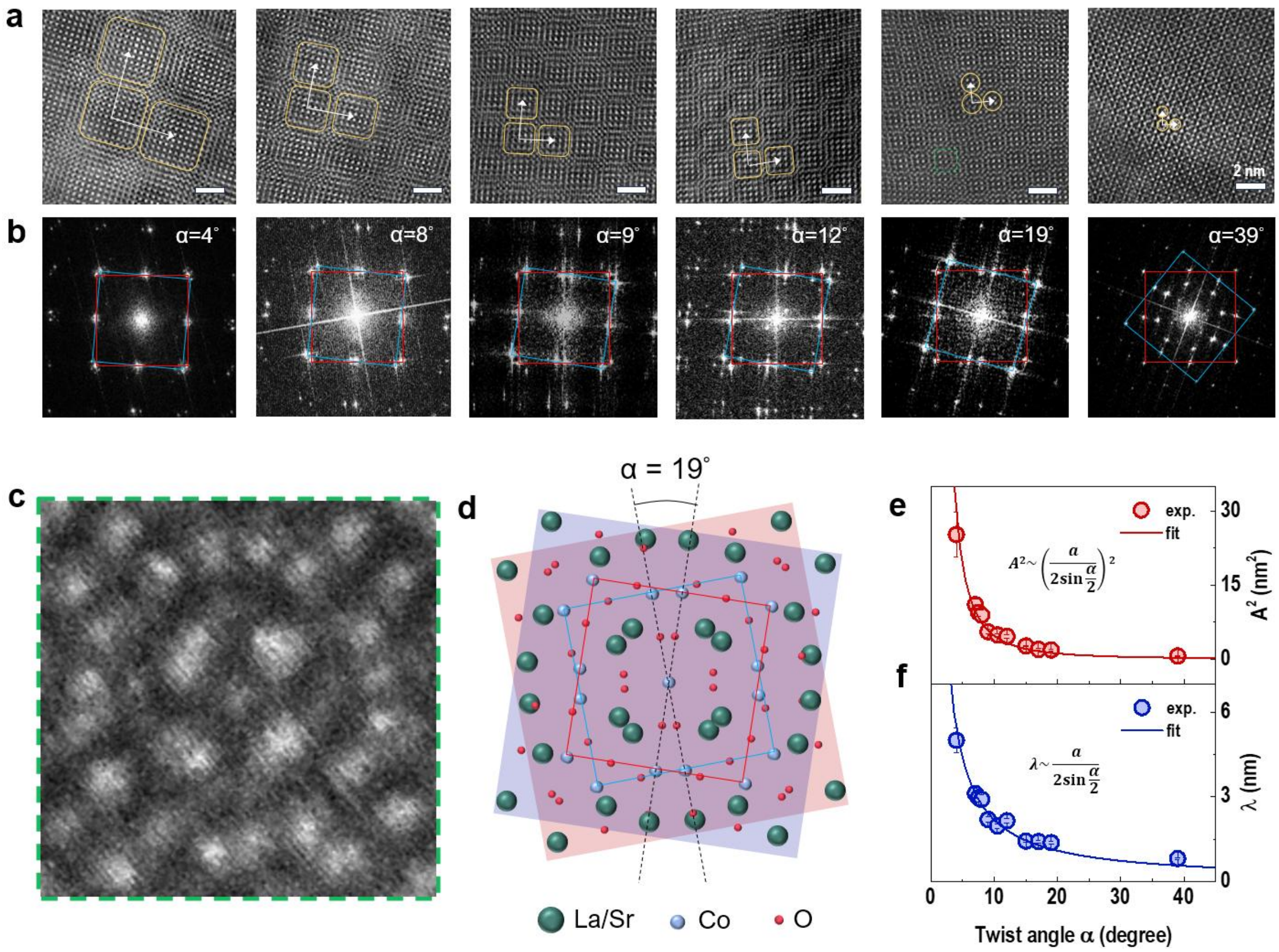

**Figure 2. Evolution of moiré patterns in twisted BL LSCO.** (a) High-resolution STEM images and (b) corresponding fast Fourier transform (FFT) images from BL LSCO membranes with various twist angles (α). Moiré patterns and their closest neighbors are indicated by yellow boxes in (a). Two distinct sets of diffraction spots, highlighted by red and blue squares, elucidate the twisted structure from BL LSCO. (c) A zoom-in STEM image from a moiré pattern with α = 19°. The atomic arrangement is illustrated in (d). (e) and (f) show the area ($A^2$) of moiré patterns and the average distance (λ) between adjacent moiré patterns as a function of α, respectively.

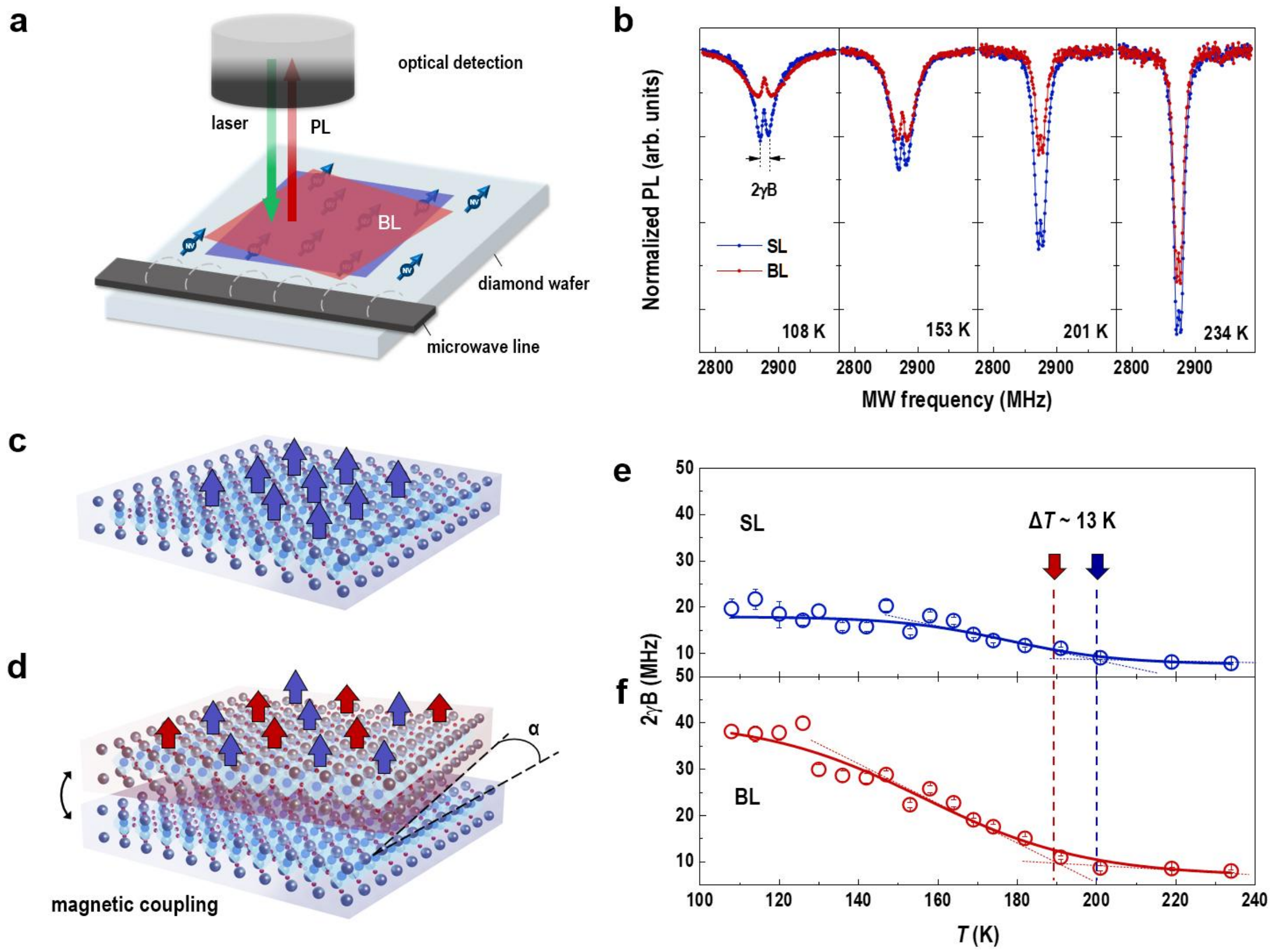

**Figure 3. Spatially resolved magnetization of twisted BL LSCO.** (a) Schematic of NV magnetometry. (b) Comparison of ODMR spectra recorded at SL and BL regions at different temperatures. All peaks split with a magnitude of $2\gamma B$, where $\gamma$=2.802 MHz/Gauss is the gyromagnetic ratio of the NV electron spin and $B$ is the local magnetic field. (c) and (d) Schematic illustrations of magnetic interactions at SL and BL regions with $\alpha = 8°$, respectively. (e) and (f) Temperature-dependent ODMR splitting at SL and BL regions, respectively. We find that the ferromagnetic-paramagnetic phase transition temperatures for SL and BL vary by approximate 10 K. Colored dashed lines are the linearly fits to the curves at FM and PM regimes.

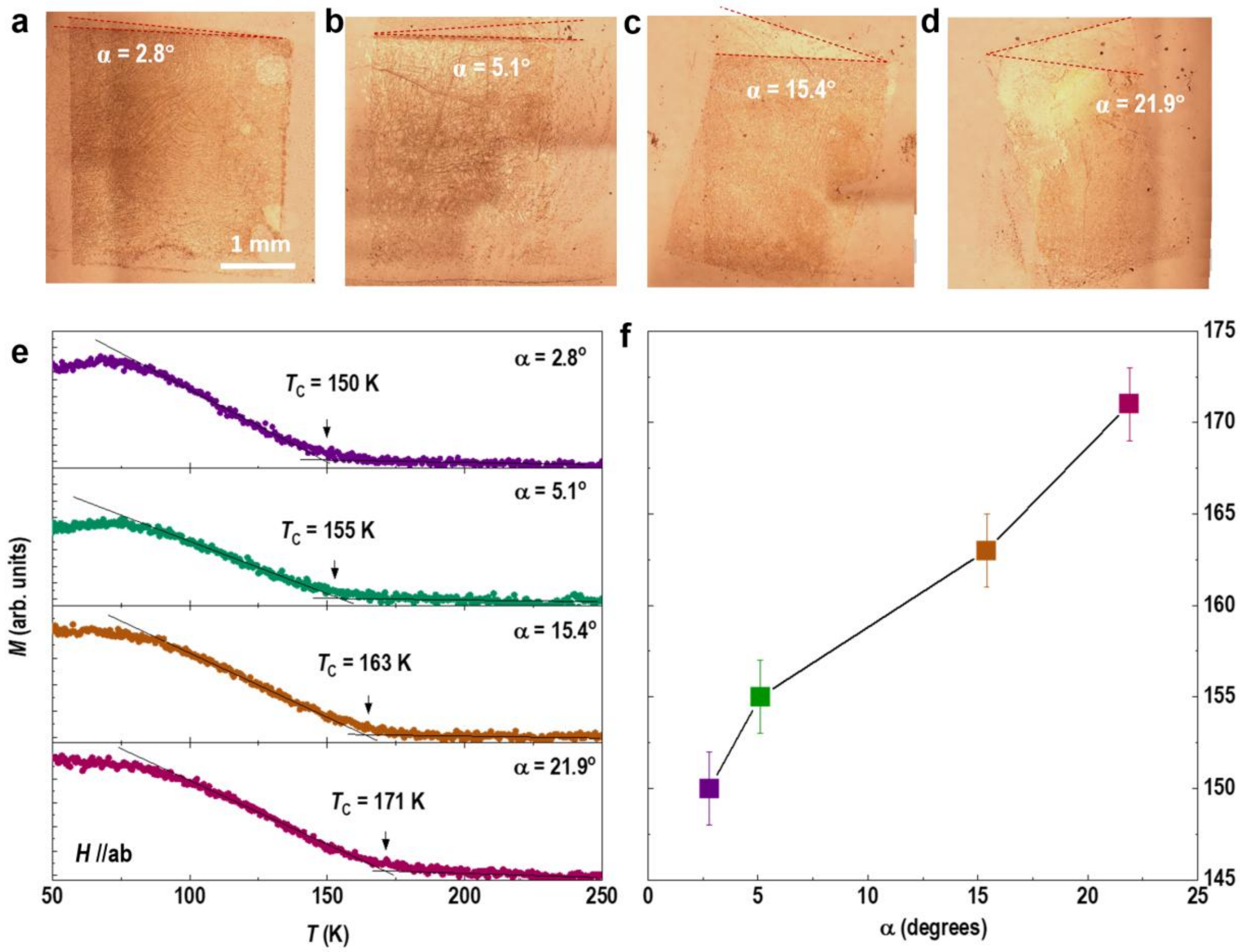

**Figure 4. Twist angle dependence of the Curie temperature ($T_C$) in BL LSCO.** (a)–(d) Optical images of twisted BL LSCO with different twist angles. (e) *M*-*T* curves of twisted BL LSCO measured along the in-plane direction during warming up at 1 kOe after field cooling. (f) $T_C$ as a function of twist angle in twisted BL LSCO.

## ASSOCIATED CONTENT

### Supporting Information

The Supporting Information is available free of charge via the internet at http://pubs.acs.org. Details on the structural characterizations, STEM sample preparation process, zoom-out STEM images, diffraction pattern simulations, temperature-dependent ODMR spectra are supplied as supporting information.

## AUTHOR INFORMATION

### Corresponding authors

*Emails: kjjin@iphy.ac.cn, dengyu@nju.edu.cn, gqliu@iphy.ac.cn, and ejguo@iphy.ac.cn

### Author contributions

◊ D. Rong, X. Chen, and S. Chen contributed equally to the manuscript. E.J.G initiated the research and supervised the project. These samples were grown by S.R.C., J.F.Z., D.K.R., H.T.H., T.C., Q.Y.W. under supervision of E.J.G. and L.F.W.; TEM lamellas were fabricated with FIB milling and TEM experiments were performed by Q.Z., Q.H.Z., and Y.D.; NV magnetometry measurements were performed by X.C., Y.X., Y.S. under supervision of G.Q.L.; D.K. and E.J.G. wrote the manuscript. All authors participated in the discussion of manuscript.

### Notes

The authors declare no competing financial interest.

## ACKNOWLEDGEMENTS

We thank the fruitful discussions with Prof. Guoqiang Yu, Prof. Yang Xu, and Prof. Wei Yang (IOP-CAS), Prof. Yu Ye (Peking University), Prof. Jinxing Zhang (Beijing Normal University), and Prof. Houbin Huang (Beijing Institute of Technology). This work was supported by the National Key Basic Research Program of China (Grant Nos. 2020YFA0309100), the Beijing Natural Science Foundation (Grant No. JQ24002), the National Natural Science Foundation of China (Grant Nos. U22A20263, 52250308, and 12304158), the CAS Project for Young Scientists in Basic Research (Grant No. YSBR-084), the China Postdoctoral Science Foundation (Grant No. 2022M723353), the Chinese Academy of Sciences (CAS) Project for Young Scientists in Basic Research (Grant No. YSBR-084), the CAS Youth Interdisciplinary Team, the Special Research assistant and Strategic Priority Research Program (B) (Grant No. XDB33030200) of CAS.

**TOC figure**

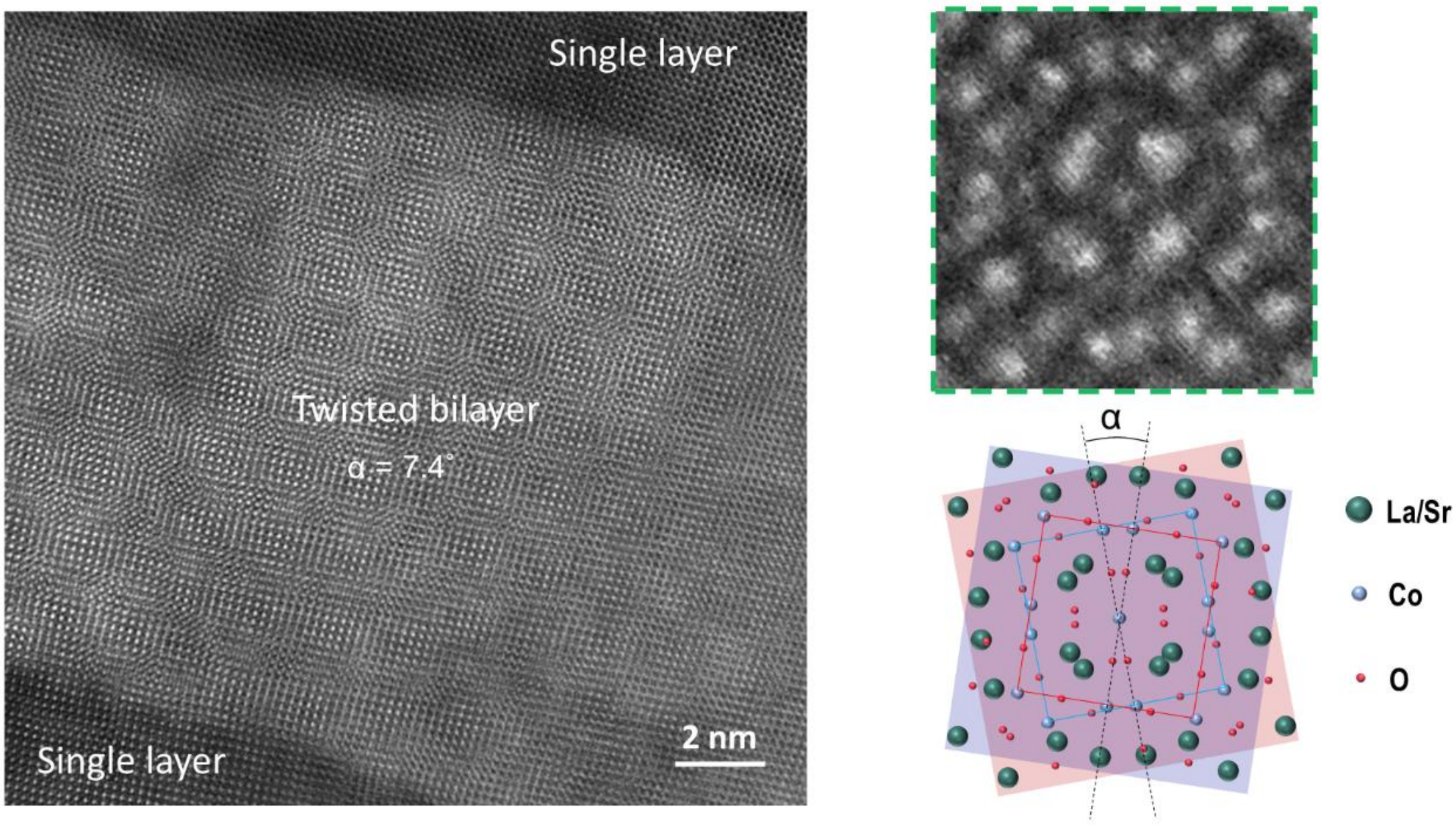
Single layer
Twisted bilayer
α = 7.4°
2 nm
Single layer
α
La/Sr
Co
O